# One4all User Representation for Recommender Systems in E-commerce

Kyuyong Shin[*,1,2], Hanock Kwak[*,1], Kyung-Min Kim[†,1,2], Minkyu Kim[1], Young-Jin Park[1,2,3], Jisu Jeong[1,2], Seungjae Jung[1,3]

[1] NAVER CLOVA, [2] NAVER AI LAB, [3] NAVER R&D Center, South Korea

{ky.shin,hanock.kwak2,kyungmin.kim.ml,min.kyu.kim,young.j.park,jisu.jeong,seung.jae.jung}@navercorp.com

## ABSTRACT

General-purpose representation learning through large-scale pre-training has shown promising results in the various machine learning fields. For an e-commerce domain, the objective of general-purpose, i.e., one for all, representations would be efficient applications for extensive downstream tasks such as user profiling, targeting, and recommendation tasks. In this paper, we systematically compare the generalizability of two learning strategies, i.e., transfer learning through the proposed model, ShopperBERT, vs. learning from scratch. ShopperBERT learns nine pretext tasks with 79.2M parameters from 0.8B user behaviors collected over two years to produce user embeddings. As a result, the MLPs that employ our embedding method outperform more complex models trained from scratch for five out of six tasks. Specifically, the pre-trained embeddings have superiority over the task-specific supervised features and the strong baselines, which learn the auxiliary dataset for the cold-start problem. We also show the computational efficiency and embedding visualization of the pre-trained features.

## CCS CONCEPTS

• **Computing methodologies** → *Neural networks*; • **Applied computing** → *E-commerce infrastructure*.

## KEYWORDS

neural networks, transfer learning, recommender system, universal user representation

---

[*]The two authors contribute equally to this work.
[†]Corresponding author.

---

## 1 INTRODUCTION

Learning general representation by pre-training large-scale models with enormous data has become increasingly common within machine learning (ML) practitioners [3, 4, 8]. This practice of training a model to perform auxiliary tasks, i.e., pre-training on pretext tasks, and then adapting the features for new target tasks, i.e., transferring to downstream tasks, is the *de facto* standard to solve a wide range of ML problems. Over the past few years, pre-trained models such as BERT [7] have achieved state-of-the-art performance in many natural language processing (NLP) and speech recognition tasks [1, 7, 14, 22]. Likewise, features of Convolutional Neural Networks pre-trained on ImageNet have made lots of progress on several computer vision (CV) tasks including object detection, semantic segmentation, and image captioning, for the last decade [25, 26]. It is empirically reported that scaling up the size of pre-training datasets and models is the key to learning highly generalizable deep features [3].

These progresses naturally raise several questions. Is the pre-training and transfer learning applicable to learning user representations for recommender systems? How much can the learned user representations cover the wide range of tasks in e-commerce platforms, from user profiling to recommendation and targeted advertising? In particular, can they enhance the performance for cold users? Is it possible to get better performance than learning from scratch? Will scaling-up the pre-training improve generalization performance?

To address these questions, we demonstrate the effectiveness of the pre-trained general user representations for recommender systems with various tasks in an e-commerce platform. Specifically, we introduce the ShopperBERT, a large model with 79.2M parameters, that learns the nine pretext tasks from the 0.8B user purchase behavior logs in the e-commerce platform. The total amount of users and items in the pre-training dataset are 12M and 48M, respectively, collected over the two years. We extract the pre-trained user representations from the ShopperBERT by mean-pooling hidden features of behavioral tokens or using the [CLS] token vector [7]. Downstream tasks for validating features pre-trained by our method include one user profiling, two targeting, and three recommendation tasks: (1) Gender Classification, (2) Membership Targeting, (3) Live Commerce Targeting, (4) Product Collection Recommendation, (5) Marketing Message Recommendation, (6) Shopping Search Query Recommendation. We evaluate our pre-training methods by comparing the performance of multi-layer perceptron (MLP), with that of the Transformers [27] trained from scratch. The MLP uses the pre-trained features as inputs while the Transformers learn from task-specific supervision signals, or, for the case of cold-start, the purchase behaviors of users in the pre-trained dataset[‡].

Our key findings for the pre-trained features are as follows:

**Generalizability**: Overall, the MLPs with pre-trained user representations show better performances than the Transformers trained from scratch for the five tasks and comparable results for the one task. We attribute this generalizability to our pre-training strategy with the nine pretext tasks and the size of pre-training dataset. We can further increase the performance by combining the pre-trained features with the Transformers.

**Benefits for cold-start problem:** The MLP obtains 4~5% of improvement over the Transformers on the two cold-start downstream tasks such as Membership Targeting and Live Commerce Targeting, in terms of AUROC, F1-score, and accuracy. The result implies that the pre-training relieves the cold-start problem by learning a more generalized user representation than the model trained from scratch, even though both use the same dataset. We also verify the performance on the other tasks by splitting the dataset into the two groups, cold and heavy user groups.

**Transfer learning can perform better than task-specific representation learning:** We show a counter-result to the previous study that task-specific representation learning with supervision has better performance over the pre-train and transfer learning [9]. Even for the downstream tasks which provide the task-specific supervisory signals, i.e., Product Collection Recommendation, Marketing Message Recommendation, and Shopping Search Query Recommendation, the pre-trained features perform better than the task-specific features. We identify that the size of the pre-trained dataset has a decisive influence on these results.

**Scale of the pre-training dataset matters:** For all tasks, the performance of pre-trained features improves as the pre-training data size increases. The result is consistent with the trend in other fields that scaling-up pre-training dataset helps improve the model's performance [12].

Our work is one of the pioneering studies addressing general user representation learning. Moreover, no previous study has comprehensively dealt with the pre-trained user representation for multiple recommender systems beyond user profiling in e-commerce to the best of our knowledge. We are planning to release our datasets and downstream tasks to the public in the form of competition to facilitate the progress of the general user representation learning. We include the discussion about the computational efficiency and embedding visualization of the pre-trained features as well as performance comparison with the fine-tuning method. Note that our goal is not just to consider alternative deep network architectures but rather to broadly explore the possibility of learning general-purpose user representation including learning strategy.

## 2 RELATED WORKS
### 2.1 Transfer Learning
Several lines of work have focused on pre-training with a large amount of data and parameters followed by fine-tuning on a specific task [7, 8]. Although there have been many reports that fine-tuning successfully improves the performance of the downstream tasks, they may suffer from a lack of extensibility. This problem is especially the case in environments where multiple pipelines operate simultaneously. When considering memory and computation time issues, having a big, ad-hoc fine-tuning model for each downstream task would be infeasible. Additionally, datasets for downstream tasks may be insufficient to tune the parameters of a huge model. For such reasons, it is increasingly popular to perform transfer learning without re-training all parameters of the pre-trained model. [4, 6, 9, 32, 33]. Recent work have showed that pre-trained language models can even perform other different tasks without fine-tuning by only demonstrating few examples [3]. We will utilize the pre-trained user representations for our downstream tasks by treating the pre-trained model as a feature extractor.

### 2.2 Pre-training Strategy for Recommendation
There have been several recent studies that use pre-trained representations for sequential recommendation. $S^3$-Rec [34] optimizes the four auxiliary self-supervised learning objectives to capture item-attribute, sequence-item, sequence-attribute, and sequence-subsequence correlations. Inspired by augmentation methods in visual representation learning, CP4Rec [30] augments user behavior sequence by cropping, masking, and re-ordering, to construct pretext tasks. Then it uses a contrastive loss function to pre-train the model. Hu et al. [11] introduced the generative pre-training framework GPT-GNN on graphs. GPT-GNN learns the inherent relationship between graph structure and node attributes through the generative process, i.e., attribute and edge generation. Note that they all fine-tune the pre-trained parameters as their focuses are not to learn general user representation [11, 30, 34]. The progresses of pre-training strategies for recommender system naturally lead to attempts to build universal user representation.

### 2.3 Universal User Representation
Deep representation learning framework to obtain universal user representations is in a beginning stage. Ni et al. [18] perform multi-task representation learning using an attention-based RNN architecture to capture in-depth representations of portal users. They assume that learning multiple tasks, e.g., predicting CTR, price preference, shop preference, etc., at once can produce better user representations. Yuan et al. [32] propose parameter-efficient transfer learning architecture named PeterRec to relieve the computational cost burden of fine-tuning. PeterRec covers five downstream tasks that include predicting user profiles like gender, age, and life status, as well as a cold-start recommendation task for browser recommender systems. Zhang et al. [33] train autoencoder-coupled Transformer networks that model retention, installation, and uninstallation collectively. They test the user embeddings in three downstream tasks for mobile app management scenarios. Gu et al. [9] propose a behavioral consistency loss to preserve the user's long-term interest and an aggregation scheme for the benefit of model capacity. The proposed method is evaluated on user preference prediction and user profiling tasks. The pre-trained models of [9, 33] play the role of feature extractor like ours.

Following the previous studies, we propose a framework to learn general-purpose user representation using self-supervised transfer learning in e-commerce platforms. Specifically, ShopperBERT pre-trains user embedding by optimizing nine pretext tasks constructed from user purchase history. This paper investigates the effectiveness of the pre-trained representation on heterogeneous e-commerce tasks including user profiling, targeting, and recommendation problems.

---
[‡]Using auxiliary data as an input for a model is a reasonable option for cold-start problems.

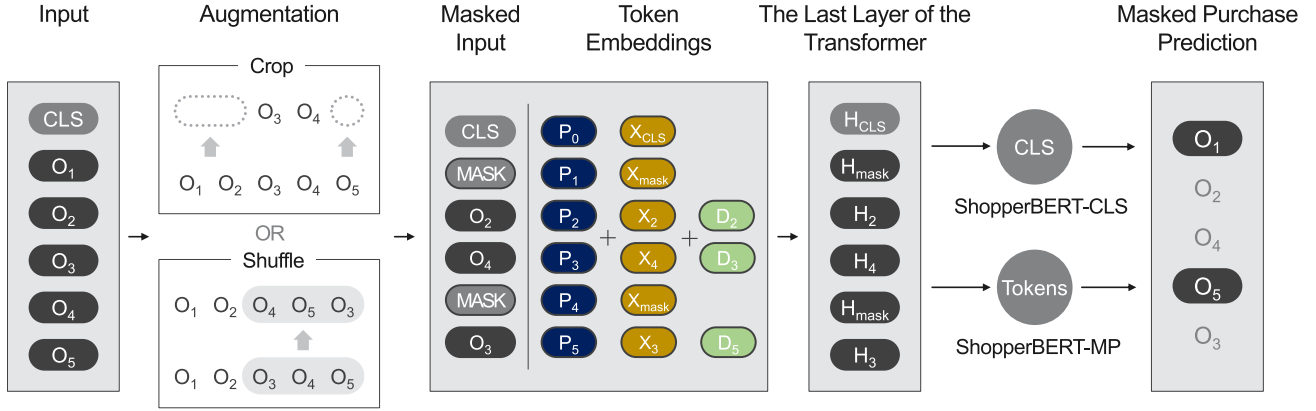

Figure 1: Overall pre-training procedure of ShopperBERT. This figure illustrates the case of shuffle augmentation. We put the [MASK] tokens after augmenting the sequence of purchase logs (O). We average the product embedding (X), date embedding (D), and positional embedding (P) of each token and pass it through the Transformer encoder. The H denotes the final hidden vector of each token, which is the last layer of the Transformer. The pretext task is to perform nine sub-tasks: predicting categories and purchase date of the masked products.

## 3 ShopperBERT

The objective of ShopperBERT is to extract a universal user representation pre-trained from large-scale user behaviors for recommender systems in an e-commerce platform.

BERT [7] is a powerful pre-training model that successfully utilizes pre-trained features to perform downstream tasks. The pretext task, Masked Language Modeling (MLM), is the key idea of BERT, which encourages the bidirectional representation of language. Recently, it has been rapidly expanded to various domains with superior performance [5, 23, 24]. As shown in Figure 1, our proposed model ShopperBERT utilizes the framework of BERT in learning general user representations, combined with the two augmentation methods.

**Model Architecture.** A user is represented by a sequence of purchase logs $s = [o_1, o_2, \ldots, o_n]$. Each purchase log contains a product feature and corresponding date. In detail, the product consists of four hierarchical categories ($c_1, c_2, c_3, c_4$) and a textual description embedding $e$ which is extracted by Sentence-BERT [22]. We use year, month, day-of-month, day-of-week, and hour to represent the date.

We need to convert the purchase log into an embedding vector to fit into the input of the Transformer. All categorical variables are initially passed into a corresponding embedding layer $g_{\mathcal{T}} : \mathcal{T} \to \mathbb{R}^H$, where $\mathcal{T}$ represents a set of categorical values. For notational simplicity, we omit $\mathcal{T}$, throughout this paper. The product embedding $x_{prod}$ is a concatenation of the categories and text feature, followed by a projection model $f_{prod}$:

$$x_{prod} = f_{prod}([g(c_1) \parallel g(c_2) \parallel g(c_3) \parallel g(c_4) \parallel f_{text}(e)]), \quad (1)$$

where $\parallel$ is the concatenation operation, and $f_{text}$ is a projection model for the textual descriptions. We use single layer MLPs for the projection models. In order to accommodate the date information, we average the $x_{prod}$ and five types of date embedding.

There are two types of pre-training methods: ShopperBERT-MP and ShopperBERT-CLS. ShopperBERT-MP performs the pretext tasks predicting masked information of the [MASK] token with the final hidden vector of that token. "MP" stands for "mean pooling" method, which averages the final hidden vectors of the behavior tokens which will be used as a user representation. On the other hand, ShopperBERT-CLS puts a [CLS] token in front of a sequence of purchase logs, and then predicts the hidden information of the [MASK] tokens by using the final hidden layer of the [CLS] token. The learned embedding of the [CLS] token is used for the user representations.

**Augmentation.** Two augmentation methods randomly transform each sequence of logs into a correlated instance. This sequence of purchase logs is transformed into different views which still implies the same hidden intention of the user's behavior. The different views with the same implication will further reinforce the robustness of user representations. There are two ways to augment the purchase logs: shuffling and cropping [30]. As pointed out in [28, 30], the user intents implied in the purchase logs are likely to be preserved even after the logs are shuffled. Thus, shuffling enhances the robustness of the model by encouraging the user representations to rely less on the order of purchase logs.

On the other hand, cropping still preserves the order of purchase logs but deletes the purchase logs that are not nearby, making the model focus more on a local view. A cropped log with a local view will focus on products that are purchased together [2], which will further train the model with relationships of these co-purchased products. Moreover, referring to [30], the model is trained to derive

**Algorithm 1** Augmentation

1: **Input:** A sequence of logs: $s$,
   A probability of cropping: $p_{crop}$
2: $l \leftarrow$ select a random integer from $[0, |s| \times 0.70)$
3: **with** probability of $p_{crop}$ **then**
4: $\quad h \leftarrow$ select a random integer from $[l + |s| \times 0.30, |s|)$
5: $\quad s_{aug} \leftarrow s[l : h]$
6: **else**
7: $\quad h \leftarrow l + |s| \times 0.30$
8: $\quad s_{aug} \leftarrow \text{copy}(s)$
9: $\quad s_{aug}[l : h] \leftarrow$ randomly shuffled $s[l : h]$
10: **Output:** An augmented sequence: $s_{aug}$

Table 1: Statistics of the Pre-training Dataset.

| Contents | Numbers |
| --- | --- |
| Users | 12,722,461 |
| Products | 48,473,095 |
| Behavior Tokens | 808,805,011 |
| Avg. Length of Sequences | 64 |
| Time Periods | 2 years (Oct, 2018 - Oct, 2020) |

Table 2: Inputs of the downstream models. The inputs are purchase logs (P), pre-trained user representations (U), and task-specific historical logs of the downstream task (T).

| Models | Pre-trained | Purchase logs | Task-specific |
| --- | --- | --- | --- |
| U-MLP | O | - | - |
| P-Trans | - | O | - |
| UP-Trans | O | O | - |
| T-Trans | - | - | O |

user representations even without sufficient information which further helps generalizing the user representation.

For all the augmentation experiments, we set all the following augmentation parameters to the same values. We augment the 30% of the data instances in the batch, and set the ratio of cropping and shuffling to 1:1 for the 30% of selected samples. When conducting the shuffling method, we randomly shuffle 30% of the purchase logs. For cropping, we crop out 0~70% of the purchase logs randomly. Algorithm 1 shows the pseudo-code of augmentation.

## 4 EXPERIMENTS

### 4.1 Pre-training

*4.1.1 Dataset.* The purchase logs imply hidden intentions. A user who purchased the iPhone is likely to purchase accessories such as iPhone cases, MagSafe, and AirPods next time. We can also assume that the user will be interested in purchasing newer technological gadgets. As such, observing a longer period of the purchase logs brings a deeper understanding of the user.

Based on the observation, we construct a pre-training dataset from our e-commerce platform. We randomly sample users who have purchase logs between Oct, 2018 and Oct, 2020. Among them, we exclude the users who purchased products less than once every two months or rank top 1% in terms of purchase frequency. As described in Section 3, purchase logs of a user consist of a date/time and corresponding product belonging to the comprehensive categories such as clothing, groceries, daily necessities, home appliances, concert tickets, etc. We do not utilize the product IDs because the product IDs of the same product may change over the date. Instead, we use the textual description of the product, along with the category information. Each product has four level of categories with vocabulary sizes of 12, 234, 2092, and 3379. We extract 768-dimensional dense representations of the description using Sentence-BERT [22]. If a user makes multiple purchases of a single product in a day, we count it as a unique purchase log. As a result, the pre-trained dataset contains 12,722,461 users, 48,473,095 products, and 808,805,011 behavior tokens collected over two years. Table 1 shows the basic statistics of the dataset.

*4.1.2 Masked Purchase Prediction.* The Masked Purchase Prediction (MPP) proceeds with nine sub-tasks, and there are task-specific prediction layers (two-layered MLP, GeLU [10]) above the Transformer outputs. These sub-tasks are to predict the four categories and five types of date information which are year, month, day-of-month, day-of-week, and hour. We randomly choose 15% of the tokens. For each chosen token, we replace it with (1) the [MASK] token 80% of the time (2) a random token 10% of the time (3) the unchanged token 10% of the time [7]. The MPP loss $\mathcal{L}_{mpp}$ is computed as the average of the cross-entropy loss $\mathcal{L}_{ce}$ of these nine sub-tasks:

$$\mathcal{L}_{mpp} = \frac{1}{|\mathcal{T}|} \sum_{\mathcal{T}} \mathcal{L}_{ce}, \quad (2)$$

where $\mathcal{T}$ is the set of nine tasks

We use the same Transformer encoder ($H$=550, $L$=20, $A$=10)[*] with max sequence length 350 for all the experiments. All the model updates use a learning rate of 0.0001 and batch size of 128. We use an Adam optimizer [13] with $\beta_1 = 0.9$, $\beta_2 = 0.999$, L2 weight decay of 0.0001, learning rate warmup over the first 30,000 steps, cosine decay of the learning rate. The total training time is 3 days. All the experiments are trained with an automatic mixed-precision package in Pytorch [19] and five V100 GPUs.

### 4.2 Downstream Tasks

*4.2.1 Downstream Models.* In the downstream tasks, we mainly evaluate the generalization ability of the pre-trained user representations with the **U-MLP** which is a simple fully connected feed-forward network (input-512-256-128-64-output, ReLU). We compare them with the two models: **T-Trans** that only uses task-specific historical logs of users, **P-Trans** that directly uses the purchase logs used in the pretext tasks, and **UP-Trans** that uses the combination of the purchase logs and pre-trained user representations to achieve better results than P-Trans. Note that using the raw purchase logs requires much more computational costs. The downstream models are summarized in Table 2. As the historical logs are time series data, we use the Transformer ($L$=4, $H$=128, $A$=4) where each token represents a single log (or event). Since the datasets of downstream tasks are relatively small and service infrastructures have limited resources, we select the relatively small model. It shares the parameters across layers which was introduced in ALBERT [14] to lower memory consumption and increase the training speed. The embedding method for the purchase log is the same as that of ShopperBERT. The Transformer has a [CLS] token to extract user embedding and we put a MLP (input-512-256-128-64-output, ReLU) on the final output of this [CLS] token to perform the downstream tasks. The UP-Trans puts the pre-trained user representations instead of the [CLS] token.

---
[*]We denote the number of Transformer blocks as $L$, the hidden size as $H$, and the number of self-attention heads as $A$. For all the Transformers, we set the hidden size of feed-forward network to be $4H$.

Table 3: Computational costs comparison of the downstream models measured from the Membership Targeting task. We measure the costs of ShopperBERT for the case of fine-tuning. The parameters include all auxiliary models such as prediction layers and embedding layers.

| Models | Inputs | Inference speedup | Parameters | Data sizes |
|---|---|---|---|---|
| ShopperBERT (fine-tuning) | Purchase logs | 1 | 79.2M | 20G |
| P-Trans | Purchase logs | 21x | 1.4M | 20G |
| U-MLP | Pre-trained user repr. | 2458x | 0.5M | 2G |

4.2.2 *Experimental Settings.* Some of the downstream tasks are recommendation tasks where the models predict the next item to recommend. In these tasks, the datasets contain positive and negative pairs $(u, i)$ of user and item. We put a positive pair $(u, i)$ if a user $u$ positively interacted with an item $i$, and the negative pairs are made by randomly sampled users and items. The goal of the training is to optimize the parameters so that the output embeddings of user and item in the positive pair are close together. We use additional models to embed items depending on the tasks. After computing the embeddings of user and item, we compute dot product between them followed by sigmoidal activation function, so the outputs are constrained to the interval (0, 1). We optimize the models with binary cross entropy loss where the labels are one and zero for positive and negative pairs respectively.

All the model updates use a batch size of 256. We use an Adam optimizer [13] with $\beta_1 = 0.9$, $\beta_2 = 0.999$, no weight decay, and exponential decay of the learning rate (decay rate = 0.995). We select the best learning rate among 0.0001, 0.0003, and 0.001 for each model. In order to utilize text information, we use Sentence-BERT [22] to extract embedding vectors from texts.

4.2.3 *Datasets.* We describe the datasets of downstream tasks in detail.

**Gender Classification (GC)**: We collect 1M users who have gender information in the e-commerce platform. We only consider the two gender types, man and woman, for this task, so it is a binary classification task.

**Membership Targeting (MT)**: Users who joined the membership have several benefits on the e-commerce platform. We can efficiently use marketing activities to engage them into the membership by identifying potential users. The task is to classify whether a user is going to join the membership in about two months. There are 951,542 users in this dataset and we set the ratio of positive and negative users to be 1:1.

**Live Commerce Targeting (LCT)**: In the live commerce service, merchants stream online videos to interact with customers and promote their products. This task is similar to Membership Targeting. The objective is to classify whether a user is going to use the Live Commerce who have not used it before. There are 1M users in this dataset and the ratio of positive and negative users is 1:1.

**Product Collection Recommendation (PCR)**: The Product Collection is a collection of products designed by merchandisers with a special category such as "Home appliances for babies", "Spring sale special offer", and "The best budget smartphones". They show a banner that is linked to a page showing the full list of products, and this task is to recommend this banner properly. We collect 1,141,244 unique click logs of these banners containing 580,700 users and 4,601 Product Collections. The recommendation models use the title of Product Collections as an item feature.

**Marketing Message Recommendation (MMR)**: Chatbot marketing is a way to engage users for the purpose of generating sales. The chatbot offers personalized events and products to users by sending interesting messages. We collect 2,401,559 click logs of 1,312,043 unique users and 20,324 messages. We use the message itself as an item feature.

**Shopping Search Query Recommendation (SSQR)**: Users can search for products from the e-commerce platform. We collect 1,584,189 search records containing 1,078,625 users and 785,476 queries. Similarly, the recommendation models use the query as an item feature.

In the Gender Classification task, there are no direct features to predict genders, so we use the purchase logs for the baseline models. The Membership Targeting and Live Commerce Targeting tasks are cold-start problems since the targeting users did not engage in the target service before. The other three tasks have their own task-specific historical logs, but some users are new or have a short history. We compare the performance of heavy and cold users in these tasks. The heavy (cold) users are top (bottom) ten percents of the users with respect to the length of task-specific historical logs.

4.2.4 *Metrics.* There are two types of the downstream tasks in the experiments: binary classification and item recommendation. We compute area under the ROC curve (AUROC), F1-score, and accuracy to evaluate the binary classification tasks. In the item recommendation tasks, we employ top-k Hit Ratio (HR@k), top-k Normalized Discounted Cumulative Gain (NDCG@k), and Mean Reciprocal Rank (MRR) by mixing the ground-truth items with 100 randomly sampled items. To test the generalization ability of the models, there are no common users between training, validation, and test sets. We repeat each experiment ten times, and then report the average value of the results.

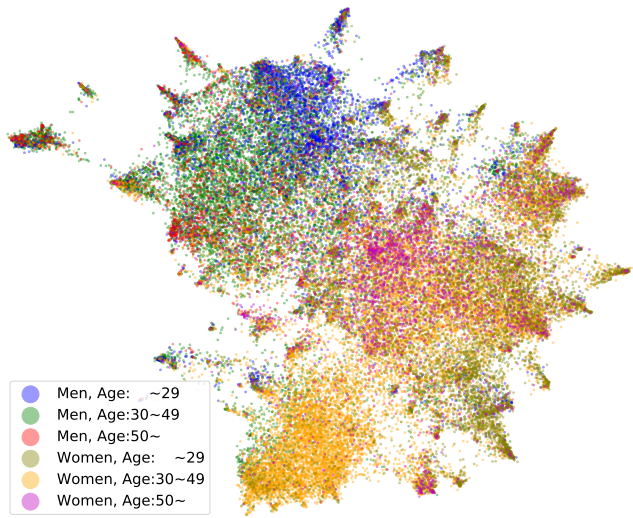

Figure 2: UMAP visualization of the user representations. The colors represent genders and ages. Note that these demographic information is not included in the pretext tasks. Best viewed in color.

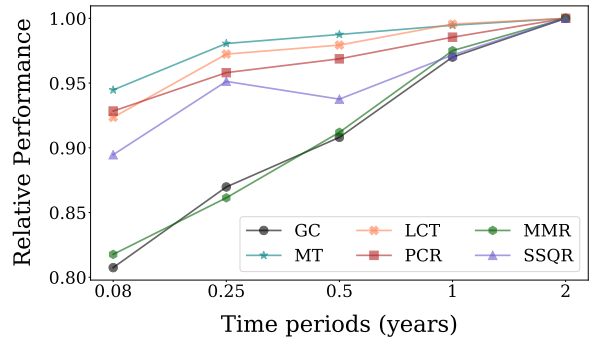

Figure 3: Comparison on the time periods of the user purchase logs used in the pretext tasks. The y-axis shows relative performance of AUROC or MRR depending on the downstream task.

## 4.3 Computational Costs

We measure computational costs of the three types of learning downstream tasks: using the pre-trained user representations with a lightweight MLP (U-MLP), using the raw purchase logs with a small Transformer (P-Trans), and fine-tuning a pre-trained Transformer (ShopperBERT). In practice, high inference speed, low storage usage, and low memory usage are critical to serving a large number of users and services. To demonstrate these goals, we measure the inference speed, the number of parameters, and the sizes of data for the Membership Targeting task. We use a single P40 GPU, disable gradient calculation, and perform 1,000 inference steps with a batch size of 128. As shown in Table 3, using the pre-trained user representations has much more advantages in terms of computational costs.

## 4.4 Visualization of the User Representations

We visualize the pre-trained user representations with respect to the demographic information of users. We randomly select 50,000 samples and use UMAP [17] for visualization, which performs general non-linear dimension reduction. Some groups of users form interesting clusters as shown in Figure 2. For instance, the two genders are divided into left and right, and the women of age more than 49 years are relatively centered. Note that this demographic information is not included in the pretext tasks. This shows that the ShopperBERT naturally learns demographic properties, implying its applicability to other user profiling tasks.

## 4.5 Effectiveness of Long Time Period in Pre-training Dataset

While many studies based on behaviorism theory have analyzed users from their behavior logs, the proper length of behavior logs is not yet determined. [15, 20, 21] assume that the user's near future behavior is strongly dependent on the last few behaviors. [2, 16, 29, 31] emphasize the importance of long-term user behaviors to learn user representations especially for recommender system. Those studies have built the training datasets with less than six months of historical logs.

In this subsection, we determine the time length suitable to learn general user representation. To validate the effect of a time length of the pre-training dataset, we conduct the ablation study. We need a longer time length than commonly used scales to deal with the problems that require behavioral logs over a year, e.g., recommending air conditioner filters, and electric devices. Specifically, we collect purchase histories of one month, three months, six months, one year, and two years of the same users. We then transfer the pre-trained features of ShopperBERT to each downstream task. As shown in Figure 3, we empirically show that ShopperBERT learns the more general user representations as the pre-training dataset has the longer time length. The result follows the trend in other domains that increasing the size of the pre-trained dataset positively affect the performance of the pre-trained model [12].

## 5 RESULTS

To validate our approach, we conduct experiments on the six downstream tasks: one user profiling, two targeting, and three recommendation tasks. We show that the simple transfer learning method outperforms the Transformers trained from scratch for the five tasks and achieves comparable results for the other empirically demonstrating the generalizability of the pre-trained features. The performances can further improve by feeding our features in the Transformer. Additionally, we verify whether our approach effectively handles cold-start scenarios on the downstream tasks. We also compare the pre-trained representations with fine-tuned ones for rich computing resource environments. The results show that the pre-trained features are competitive given that the performance

Table 4: Results on the downstream tasks. We compare the pre-trained representations (U-MLP) with the baselines learned from the raw purchase logs (P-Trans), the combination of them (UP-Trans), and the task-specific supervised representations (T-Trans) in terms of the three evaluation metrics. "+A" means the augmentation. The results are averaged over ten repeated experiments. The best results among U-MLPs and all models are denoted in bold and underlined fonts, respectively.

| Downstream tasks | Metrics | U-MLP | | | | P-Trans | UP-Trans | T-Trans |
|---|---|---|---|---|---|---|---|---|
| | | MP | MP + A | CLS | CLS + A | - | MP | - |
| Gender Classification (GC) | AUROC | 0.9420 | 0.9426 | 0.9453 | **0.9480** | 0.9669 | <u>0.9670</u> | - |
| | F1-Score | 0.8937 | 0.8939 | 0.8978 | **0.9009** | <u>0.9267</u> | 0.9260 | |
| | Accuracy | 0.8749 | 0.8753 | 0.8799 | **0.8837** | <u>0.9140</u> | 0.9135 | |
| Membership Targeting (MT) | AUROC | 0.6744 | **0.6766** | 0.6245 | 0.6290 | 0.6412 | <u>0.6860</u> | - |
| | F1-Score | 0.6463 | **0.6473** | 0.6039 | 0.6061 | 0.6111 | <u>0.6476</u> | |
| | Accuracy | 0.6271 | **0.6280** | 0.5912 | 0.5946 | 0.6027 | <u>0.6351</u> | |
| Live Commerce Targeting (LCT) | AUROC | 0.7018 | **0.7039** | 0.6551 | 0.6595 | 0.6654 | <u>0.7073</u> | - |
| | F1-Score | **0.6566** | 0.6555 | 0.6132 | 0.6158 | 0.6262 | <u>0.6599</u> | |
| | Accuracy | 0.6464 | **0.6479** | 0.6121 | 0.6159 | 0.6202 | <u>0.6505</u> | |
| Product Collection Recommendation (PCR) | HR@10 | 0.1425 | 0.1425 | 0.1425 | **0.1426** | 0.1424 | <u>0.1427</u> | 0.1418 |
| | NDCG@10 | 0.7837 | 0.7832 | 0.7833 | **0.7841** | 0.7804 | <u>0.7858</u> | 0.7635 |
| | MRR | 0.6938 | 0.6932 | 0.6934 | **0.6947** | 0.6887 | <u>0.6964</u> | 0.6742 |
| Marketing Message Recommendation (MMR) | HR@10 | 0.1318 | 0.1324 | 0.1332 | <u>**0.1336**</u> | 0.1149 | 0.1267 | 0.1146 |
| | NDCG@10 | 0.5825 | 0.5868 | 0.5923 | <u>**0.5944**</u> | 0.4956 | 0.5578 | 0.4945 |
| | MRR | 0.4486 | 0.4527 | 0.4589 | <u>**0.4606**</u> | 0.3610 | 0.4245 | 0.3601 |
| Shopping Search Query Recommedation (SSQR) | HR@10 | 0.0543 | 0.0546 | 0.0552 | <u>**0.0555**</u> | 0.0532 | 0.0544 | 0.0544 |
| | NDCG@10 | 0.2827 | 0.2840 | 0.2861 | <u>**0.2871**</u> | 0.2769 | 0.2814 | 0.2832 |
| | MRR | 0.2259 | 0.2267 | 0.2286 | <u>**0.2294**</u> | 0.2212 | 0.2251 | 0.2275 |

Table 5: Performance comparison of fine-tuning the pre-trained ShopperBERT-MP model and U-MLP (MP+A) on the two targeting tasks.

| Tasks | Metrics | Fine-Tuning | U-MLP |
|---|---|---|---|
| MT | AUROC | **0.6775** | 0.6766 |
| | F1-Score | 0.6432 | **0.6473** |
| | Accuracy | **0.6295** | 0.6280 |
| LCT | AUROC | **0.7070** | 0.7039 |
| | F1-Score | 0.6528 | **0.6555** |
| | Accuracy | **0.6500** | 0.6479 |

gaps are small despite the large resource consumption difference as shown in Table 3.

## 5.1 Generalizability of the User Representations

We present the generalizability of the pre-trained user representations from ShopperBERT on the six downstream tasks. Results are presented in Table 4.

**Targeting.** The best method, MP+A, significantly surpasses the P-Trans in the targeting tasks in terms of AUROC. The MP+A achieves 0.6766 and 0.7039 AUROC scores on Mebership Targeting and Live Commerce Targeting, respectively, increased by 5.5% and 5.8% compared to the P-Trans. The UP-Trans outperforms the others obtaining 6.0% AUROC score improvement over the P-Trans model. Other metrics such as F1-Score and Accuracy follow similar trends.

We also compare the U-MLP, which uses transfer learning by feature extraction, with the fine-tuned method. Table 5 shows performance comparisons on the targeting tasks. As a result, the feature extraction method degrades the performances in terms of AUROC by only 0.13% and 0.43%, respectively.

**Recommendation.** For Marketing Message Recommendation and Shopping Search Query Recommendation, all methods of U-MLP outperform the P-Trans on both tasks by a substantial margin. The best methods CLS+A obtains 19.9% and 3.6% NDCG@10 score improvements over the P-Trans, respectively. Interestingly, the performance of UP-Trans is inferior to that of the U-MLP, but still outperforms the P-Trans by 12% and 1.6% in terms of NDCG@10.

For Product Collection Recommendation, the experimental results of CLS+A are also higher than that of the P-Trans. Unlike Marketing Message Recommendation and Shopping Search Query Recommendation, the results of UP-Trans show performance improvement over the P-Trans.

**User Profiling.** Out of all the downstream tasks, Gender Classification is the only one that predicts user demographic information. The U-MLP and UP-Trans perform competitively with the P-Trans. We speculte that the gender classification task is so simple that our pretext task cannot provide any additional help in the training process. However, as indicated in Table 3, the U-MLPs are generally

Table 6: Performance comparison of T-Trans and U-MLP (CLS+A) for cold (bottom 10%) and heavy (top 10%) users. The results are averaged over ten repeated experiments.

| Tasks | Cold/Heavy | Metrics | T-Trans | U-MLP |
|---|---|---|---|---|
| PCR | Cold | HR@10 | 0.0924 | **0.0932** |
| | | NDCG@10 | 0.7276 | **0.7546** |
| | | MRR | 0.6671 | **0.6995** |
| | Heavy | HR@10 | 0.3799 | **0.3804** |
| | | NDCG@10 | 0.8812 | **0.8875** |
| | | MRR | 0.6673 | **0.6729** |
| MMR | Cold | HR@10 | 0.0661 | **0.0770** |
| | | NDCG@10 | 0.4129 | **0.5204** |
| | | MRR | 0.3522 | **0.4536** |
| | Heavy | HR@10 | 0.3434 | **0.3998** |
| | | NDCG@10 | 0.7637 | **0.8300** |
| | | MRR | 0.3627 | **0.4624** |
| SSQR | Cold | HR@10 | 0.0368 | **0.0378** |
| | | NDCG@10 | 0.2422 | **0.2453** |
| | | MRR | 0.2232 | **0.2254** |
| | Heavy | HR@10 | **0.1428** | 0.1415 |
| | | NDCG@10 | **0.4658** | 0.4652 |
| | | MRR | **0.2368** | 0.2355 |

faster than the others, the trade-off against performance is still worth considering.

Overall, the models with the augmentation show outstanding performances in all cases except one task. It shows that the augmentation methods help to learn general-purpose user representation by playing a role of regularizer.

### 5.2 Performance on Cold-Start Problem

As described in Section 4.2.3, Membership Targeting and Live Commerce Targeting tasks have the complete cold-start scenarios, where the users have no prior engagement. The performance gain of the U-MLP over the T-Trans imply that even though the models use the same datasets, our pre-training framework handles the cold-start problem better by learning a more generalized user representation.

For an in-depth verification of the advantage in the cold-start scenarios, we extend our experiments to the recommendation tasks. We split the users into two groups, cold and heavy users, based on the engagement frequency for each service. The cold user group corresponds to the bottom 10% while the heavy group represents the top 10%. We use the T-Trans as a baseline because the behavior logs are available for both groups. Table 6 shows the performance comparisons between T-Trans and U-MLP for the recommendation tasks. For cold user group, the U-MLP obtains 3.7%, 26.0% and 1.2% NDCG@10 score improvement over the T-Trans for PCR, MMR, and SSQR, respectively. Overall, the U-MLP achieves the best performances on all metrics. The following subsection describes the performance comparison with T-Trans in the heavy groups.

### 5.3 Comparison with the Task-specific Supervised Representations

Previous studies argue that task-specific representation learning with the supervision signals has better performance over the pre-training and transfer learning [9]. However, we show the counter results. As shown in Table 4, our best method, CLS+A, comfortably outperforms the T-Trans in PCR, MMR, and SSQR in all metrics. We attribute these counter results to the size of the pre-training dataset. Figure 3 shows that the U-MLPs fail to perform better than the T-Trans if the time length of the historical logs is relatively short: one to six months. Increasing the time length of pre-training dataset leads to a strict performance improvement, which eventually outperforms the T-Trans across the three tasks. The results agree with [12] proving that increasing the size of pre-training dataset reduces the errors of the models.

It is also surprising that our pre-training framework is able to achieve better or comparable results than the T-Trans on the heavy user groups. As presented in Table 6, the U-MLP achieves 0.8875 and 0.8300 NDCG@10 scores on Product Collection Recommendation and Marketing Message Recommendation, respectively, obtaining 0.7% and 8.7% improvements over the T-Trans. The U-MLP performs comparably with the T-Trans for Shopping Search Query Recommendation.

## 6 CONCLUSION AND DISCUSSION

We pre-train a large-scale model with the long-term real-world data to learn general-purpose user representations, and then successfully apply to the heterogeneous end tasks with the lightweight models achieving the promising results. The U-MLP outperforms the T-Trans and P-Trans in the targeting and recommendation tasks. It is also much more cost-effective than fine-tuning so that it is more likely to satisfy system requirements. We further study the augmentation methods and structures of ShopperBERT, and the hybrid methods to enhance the generalization ability of the user representations. We also show its applicability to other user profiling tasks through visualization of the learned embeddings.

There is no superior learning method between MP and CLS in the results. It would be possible to design more effective learning methods in large-scale data. The user representation of historical logs is similar to the sentence embedding of language models. The Sentence-BERT [22] uses labels of sentence pair, e.g., entailment, contradiction, and neutral, for the siamese network structure, but there are no general labels for user pairs. Therefore, we need specialized methods for learning general-purpose user representations. A more careful study of this subject is left for future research.

The research progress in this field is relatively slow due to the absence of public datasets collected from real-world platforms which explicitly distinguishes pre-training and downstream data. The previous works have combined public recommendation datasets or used their private data. We are planning to release our datasets and downstream tasks to the public in the form of competition to find more superior methods.

In the future, we will extend the coverage of our user representations to other domains such as movies and news by extending pre-training datasets.